# Inhomogeneous Gain Saturation in EDF: Experiment and Modeling

Romain Peretti, Bernard Jacquier, David. Boivin, Ekaterina Burov, Anne-Marie. Jurdyc

*Abstract*—Erbium-Doped Fiber Amplifiers can present holes in spectral gain in Wavelength Division Multiplexing operation. The origin of this inhomogeneous saturation behavior is still a subject of controversy. In this paper we present both an experimental methods and a gain's model.  Our experimental method allow us to measure the first homogeneous linewidth of the 1.5 μm erbium emission with gain spectral hole burning consistently with the other measurement in the literature and the model explains the differences observed in literature between GSHB and other measurement methods.

## I. Introduction

Since optical maser [1] and laser [2] were demonstrated, the need for modeling amplification by stimulated emission became important [3]. More than 20 years after, the first optical amplification in Erbium-Doped Fiber Amplifier (EDFA) was achieved and reported [4]. Since that time, Wavelength Division Multiplexing (WDM) boosted the already huge capacity of the fibers, taking advantage of the broad spectral gain of the EDFA. Models based on the homogenous spectral gain profile were very popular but they do not work for multi-wavelength operation [5].

Actually, limitations of gain performance, spectral bandwidth or signal saturation of the gain [6] are still controversial questions which emphasize a number of new developments such as chemical composition of glass materials and engineering of the micro and nanostructuration of the fiber. The nature of the broadening of the $^4I_{15/2} \leftrightarrow {}^4I_{13/2}$ transition in the near infrared has been the subject of several investigations using various laser spectroscopy techniques in different ranges of temperature. It does not allow pointing out real parameters to fully describe optical performances of EDFAs and usable for material optimization.

On the one hand, using a saturation spectroscopy technique (Gain Spectral Hole Burning,(GSHB)), hole width and depth have been already reported in the late nineties [7,8] showing that they change with signal wavelength and pump power but differently with glass composition from silica to fluoride [9] and even for other rare earths [10].

On the other hand, a Resonant Fluorescence Line Narrowing (RFLN) technique [11] was used for line width measurements in a number of bulk glasses available for fiber devices [12]. These authors point out some discrepancy between their results and GSHB earlier measurements reported by Desurvire *et al*. [7]. In our previous work [13], we gave first the results of both experiments useful to compare the two experimental approaches, but some questions were still opened.

A full complex model is needed to reproduce first all the previous experiments and behaviors of EDFA, and second to suggest an understanding and interpretation of "secondary hole" or "anti-hole" observed in literature [14, 15]. In this report, we establish the first step of a model taking into account the inhomogeneous spectral profile in EDFA to better understand GSHB and RFLN observations.

## II. Experiments

RFLN and GSHB experiments were carried out within the same conditions including a well suited temperature. The homogeneous line width of erbium in glass is measurable by the RFLN technique within the limits of our instrumentation between 30 K and 120 K, GSHB can be detected between 30 K and 300 K. However, at temperatures greater than 120 K, the hole width is too broad to be quantitatively measured. So, for these reasons and experimental simplicity, liquid nitrogen temperature (77 K) was chosen for both experiments. The investigated sample is a standard EDF from Draka, i.e. 0.04 wt% erbium-doped aluminosilicate fiber.

Romain Peretti, Bernard Jacquier and Anne-Marie. Jurdyc are with Université de Lyon, Université Lyon 1, CNRS/LPCML, Villeurbanne 69622 France. (corresponding author Anne-Marie. Jurdyc phone: +33 (0)472 431 41 ; fax: +33(0)4 724 311 30; e-mail: anne-marie.jurdyc@univ-lyon1.fr).

David. Boivin, and Ekaterina. Burov are with Draka, Draka Communication, route de Nozay, Marcoussis, 91460, France.

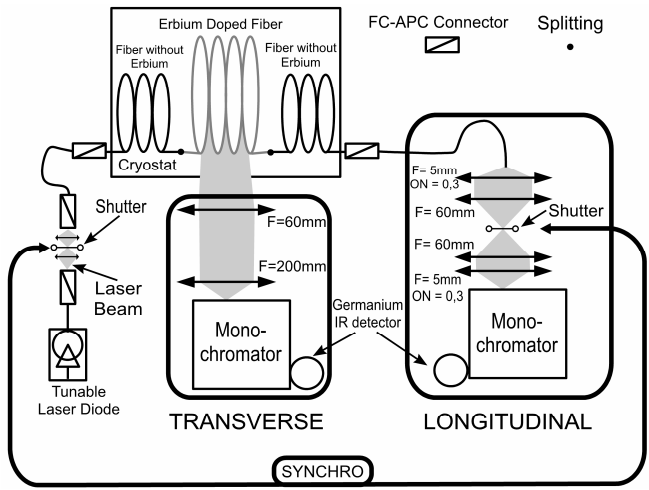

Fig. 1: Experimental set-up used for RFLN measurements.

*A. RFLN Experimental set-up*

Fig. 1 represents a schematic of the experimental set-up of RFLN. Two detection channels were used: the non-propagating transverse light into the fiber (unguided spontaneous emission flux) and the propagating light detected directly through the output of the fiber. In the following, the first channel is called transverse RFLN and the second one, longitudinal RFLN. In both cases the fiber is excited using a narrow and tunable infrared laser diode with a typical line width of 100 MHz over the 1480–1580 nm wavelength range during the wavelength scan (typically few minutes), due to the time jitter. The output laser power is a few milliwatts. The beam is focused on a mechanical shutter used to modulate the amplitude of the signal, and then we inject it in a SMF-28 fiber. Losses through objectives and shutter and fiber re-injection reduce the laser power to a maximum of 500 µW measured at the output of this fiber. To be fully immersed in the cryostat containing liquid nitrogen (77 K) the EDF under test is spliced to SMF-28 pigtail witch prevents back reflection from interfering with the excitation signal. An FC-APC connector allows the pump to be injected into the system.

Then, for transverse RFLN, since the cryostat is equipped with optical windows, emitted light is collected by a lens and focused to the entrance slit of a Jobin-Yvon U1000-IR double monochromator. The signal is detected by a slow but high-sensitivity germanium-cooled IR detector from North Coast. In that configuration, laser scattered light from the fiber is measured rather weak and does not perturb the response of the germanium detector. Then, spectral discrimination is used to distinguish fluorescence from the laser scattered light. A digital lock-in amplifier allows amplifying and recording the signal.

For the longitudinal RFLN, light is collected by an objective from the output of the fiber, focused on a second mechanical shutter, and then sent at the entrance slit of the same monochromator and detector as above. In that configuration, the excitation laser is directly sent in our detection system and strongly perturbs the germanium detector. That is why we place a second mechanical shutter, isochronous and out of phase with the first one to avoid blinding the detector. The fastest response of the two shutter system we measured was about 100 µs, that permits to measure fluorescence at the excitation wavelength [12]. Again, a digital lock-in amplifier is used to amplify the signal. We repeated the experiments several times in order to get the order of magnitude of the precision of our measurements.

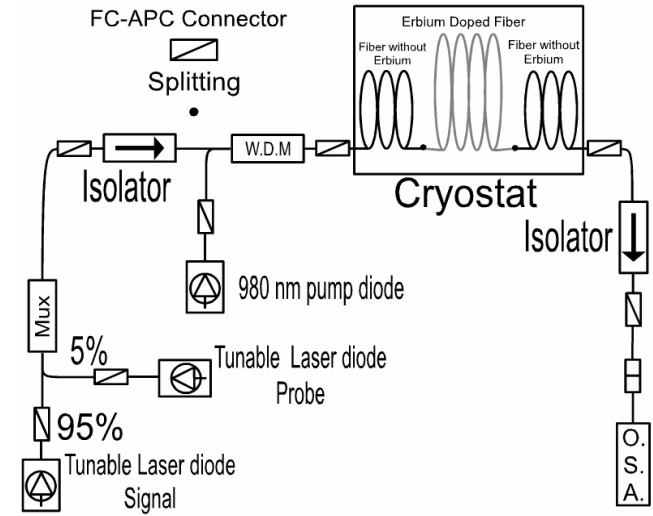

Fig. 2: Experimental set-up used for GSHB measurements.

*B. GSHB experimental set-up*

Fig. 2 shows a schematic of the experimental set-up of GSHB. This experiment is performed with CW lasers without any modulation. The set-up is basically a fiber amplifier with a 200 mW single mode laser diode which pumps the erbium ions at 980 nm. The signal beam used to burn the hole is the same used in the RFLN experiment, a 10 mW maximum tunable laser, with a 100 MHz line width limited by the time jittering. A similar laser diode provides the probe beam but with a maximum power less than 1mW. An attenuator allows decreasing the power without perturbing the wavelength and the beam quality. Both 1.55 µm signals and probe beams were multiplexed, going through an isolator to be wavelength multiplexed with the pump beam. All three lasers go through a SMF-28 fiber spliced to our erbium-doped fiber immersed into the liquid nitrogen cryostat (77 K). After that, a second isolator prevents the erbium-doped fiber to be parasited by laser effect, and the probe signal is analyzed with an Anritsu MS9710B optical spectrum analyzer (O.S.A.).

The experiment follows the typical procedure. In the first step, the wavelength and power of the signal beam are fixed to burn a hole in the spectral gain. The power of the pump beam is also set to get the desired population inversion in the EDFA. In the second step, we choose the right power of the probe beam to avoid any further saturation effect. Finally the wavelength of the probe beam is tuned around the burning signal wavelength in order to get the gain spectrum for these wavelengths. The gain is measured using the "max-hold" function of the O.S.A. We repeated the experiments several times in order to get the order of magnitude of the precision of our measurements.

## III. Results

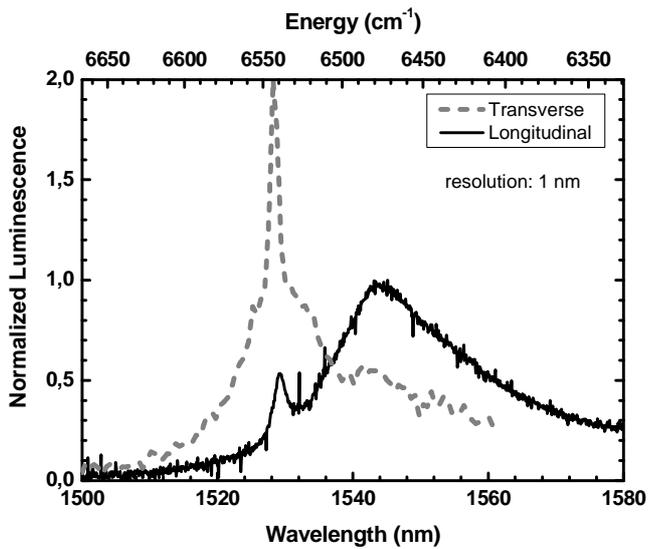

Fig. 3: Comparison of transverse and longitudinal RFLN at 77 K under 1528 nm excitation (power 0.5 mW) for a 6 m long fiber.

### A. RFLN results

Fig. 3 shows a comparison of transverse and longitudinal RFLN measurements at 77 K under 1528 nm excitation (power 0.5 mW) for a 6 m long fiber. As seen in Fig. 3 the transverse RFLN spectrum reproduces the spontaneous emission rate at temperature of analysis. At 1528 nm the sharp line is the laser light. Since our resolution is limited to 1nm because of the strength of the fluorescence signal we cannot spectrally discriminate the laser from the resonant fluorescence. The longitudinal RFLN signal results from a well known amplified spontaneous emission (ASE) process in the erbium-doped fiber together with the resonant fluorescence at 1528 nm. This observation implies that transverse RFLN is a more direct measurement than the longitudinal one. But since the signal in transverse RFLN detection mode is too weak to be exploited, the following part will concern longitudinal RFLN results.

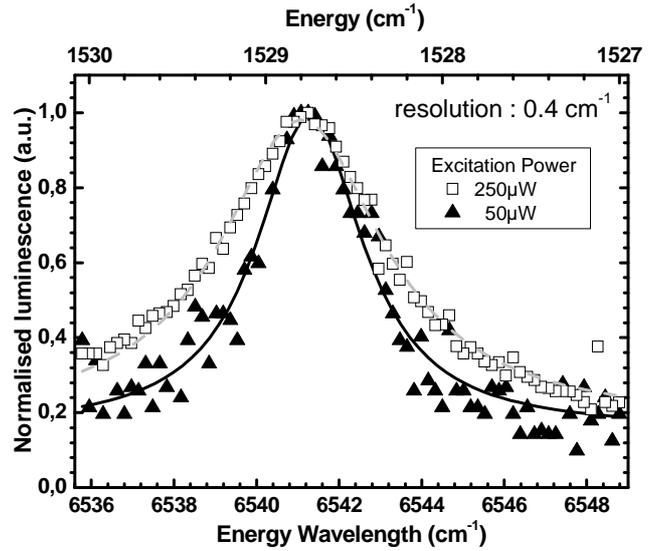

Fig. 4: Measured longitudinal RFLN at 77 K for the fiber length of 6 m (symbol) and Lorentz fit (line) at 1528 nm for the laser power: 250 μW (open squares and a gray dash line) and 50 μW (black triangles with a black solid line).

In Fig. 4, we present longitudinal RFLN line width measurements and fittings with a Lorentz line profile at 1528 nm for two pump powers: 250 μW and 50 μW. All the data have been recorded at the temperature of 77 K for a 6 m long fiber. The two spectral distributions show a nice lorentzian profile. Such a profile is characteristic of a homogeneous line width behavior (because a Lorentz shape is the Fourier transform of single-exponential time decay). The FWHM is extracted from this fit so that it can be easily analyzed as a function of different parameters such as: excitation pump power, fiber length, and excitation wavelength.

Fig. 5 shows the dependence of FWHM line width (deduced from the lorentzian fit) as a function of laser excitation power set at two different wavelengths with a 0.4 cm$^{-1}$ (0.1 nm) spectral resolution

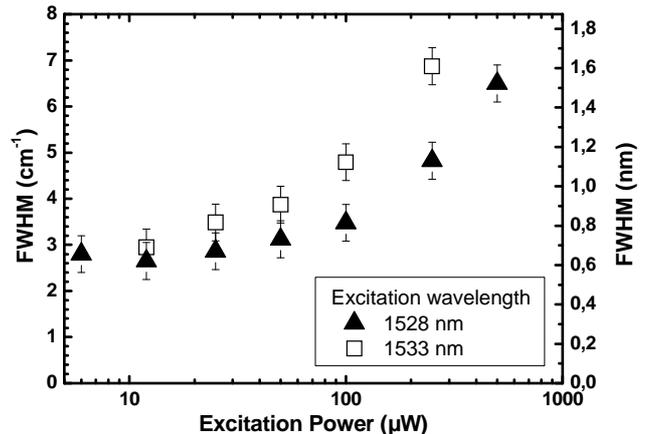

Fig. 5: FWHM of RFLN line width (0.4 cm$^{-1}$ resolution) from Lorentz fit in function of excitation power for two wavelengths: 1535 nm (open squares) and 1530 nm (black triangles) All the data are recorded at 77 K for a 6 m long fiber.

A strong power dependence of the width can be observed from 10 µW and toward 1mW. However, at low power the FWHM reaches an asymptotic value of 2.8 +/- 0.2 cm$^{-1}$ (0.65 +/- 0.05 nm) independently of the analyzed wavelength in the 1530 nm window.

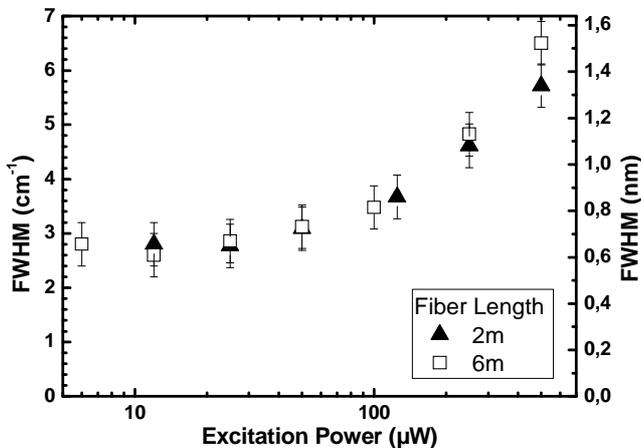

Fig. 6: FWHM of RFLN line width (0.4 cm$^{-1}$ resolution) from the Lorentz fit versus excitation power for two lengths of the fiber: 6 m long (open squares) and 2 m long (black triangles. All the data are recorded at 77 K.

Fig. 6 presents the dependence of the FWHM of RFLN line width (0.4 cm$^{-1}$ / 0.1 nm resolution) from Lorentz fit as a function of pump power for two lengths of the fiber. A similar behavior is obtained for the two lengths until the pump power reaches 500 µW.

It is noted that the two dependences (Figs. 5 and 6) show the same asymptotic value quoted above.

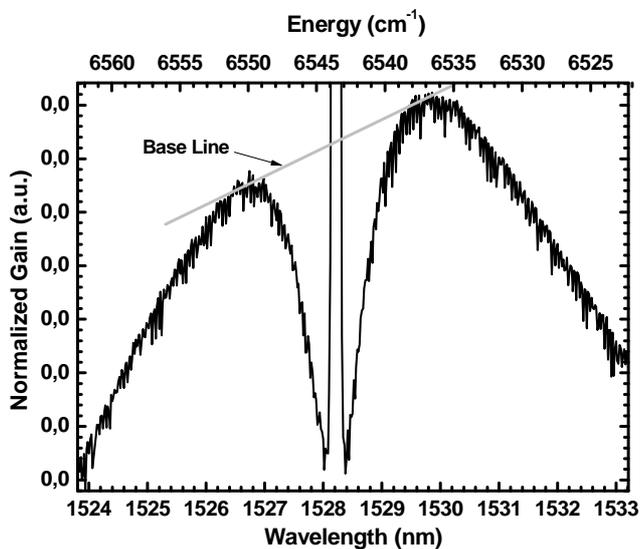

Fig. 7: GSHB measurement at 120 mW pump power, for a burning signal of 200 µW at 1528 nm, at 77 K with a spectral resolution of 0.07 nm.

*B. GSHB results*

A typical GSHB result is shown in Fig. 7, for a burning signal of 200 µW at 1528 nm at 77 K with a resolution of 0.07 nm (0.3 cm$^{-1}$) and a pump at 980 nm (10204 cm$^{-1}$) power of 120 mW. To extract a parameter from these data by fitting the hole with a line profile is not easy due to the lack of knowledge of the zero origin of the amplitude. Usually, in literature, the hole width is extracted from the difference of two gain spectra measured in an hypothetical same inversion but different burning conditions. In fact, in an inhomogeneous manifolds system, this hypothesis is not realistic. That is the reason we decided to basically measure the FWHM just after removing the bitangential baseline (the straight line that is the tangent of the curve on 2 points, here on each hump around 1526 and 1531 nm) i.e. the broad absorption spectrum (see Fig. 7). This method is valid because the ratio of the hole width to the inhomogeneous linewidth is small at 77K, but cannot be used at room temperature. With this method the hole width has been depicted in Fig. 8 as a function of the logarithm of signal power.

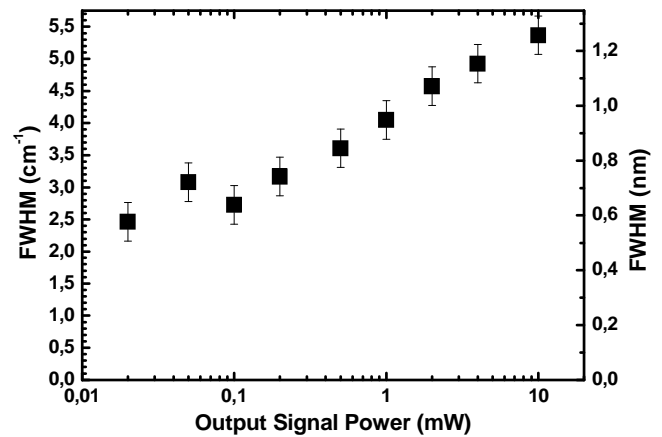

Fig. 8: GSHB width as a function of output power of a 6 m long fiber at 77 K.

There is a quasi-exponential dependence of the width between 0.1 mW and 10 mW, but at low power, the hole width reaches an asymptotic average value of 2.8 cm$^{-1}$ (0.65 nm) similar to the value obtained by RFLN experiments.

IV. MODELING

From the results reported above in Fig.7, we demonstrated that GSHB effect should be considered as a consequence of inhomogeneous broadening of the spectral gain of EDF. It is confirmed by the RFLN experiments that indicate the nature of the inhomogeneous broadening recorded in the two experiments at 77 K. To develop our model of amplification of light, in this work, we first did not take into account its propagation along the fiber. This is rather consistent because we do not observe any significant effect arising from the change in the fiber length as shown in Fig. 6 for the short fiber length. This propagation will be considered in a further work for modeling the behavior in real EDFAs. However, to improve the existing models, we do take into account the random distribution of the erbium ions in different crystalline environments of the glass net that gives rise to the inhomogeneous spectral line widths of the resonant and non-resonant electronic lines.

Despite Stark splitting making sublevels in manifolds and then Er$^{3+}$ electronic structure in glass a "21 levels system" (6 for the $^4I_{11/2}$ manifold, 7 for the $^4I_{13/2}$ manifold, and 8 for the $^4I_{15/2}$ manifold), we are using the usual 3 levels system theory to represent one erbium site because, in silicate materials, phonon relaxation of upper excited states is very fast [16]. Moreover, we have neglected any up-conversion effect [17, 18]. The inhomogeneous behavior is introduced by considering hundreds of sites in our simulation, depicting the inhomogeneity of spectral properties of erbium ions population. Because our modulation is bellow 10 kHz in our measurement and of future upgrade of our model (like up-conversion and energy transfer), we choose to solve the system in the stationary conditions. Notice that this assumption gives a correct answer for modulation up to 10 kHz [19, 20, 21].

A schematic of our model for one site is shown in Fig. 99.

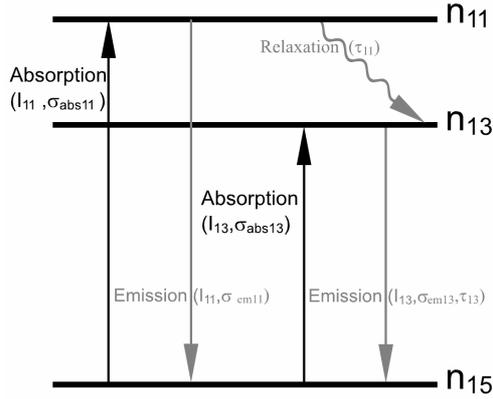

Fig. 9: 3 levels system of one erbium site.

Here, $n_d$ is the population of the $^4I_{d/2}$, $\tau_d$ is the lifetime of the level $^4I_{d/2}$, $\sigma_{absd}$ and $\sigma_{emd}$ are the absorption and emission cross sections of the level $^4I_{d/2}$, $\sigma_{absd}$, $I_d$ is the intensity of the light at a wavelength of the d level energy $\nu_{11}$ (~980 nm (10204 cm$^{-1}$) and ~1530 nm (6535 cm$^{-1}$) for the $^4I_{11/2}$ and $^4I_{13/2}$ respectively.

The electronic transition $^4I_{13/2} \rightarrow ^4I_{15/2}$ is supposed to be purely radiative, and any inhomogeneous effect of the $^4I_{11/2}$ level is neglected. Actually the pump beam is assumed to be broad in order that all erbium's sites absorb the same power. In this way, we will not depict any pump-mediated inhomogeneity [22].

Taking the following intermediate variables presented in (1), into account,

$$R_{11} = \frac{I_{11}}{h\nu_{11}}\sigma_{abs11} \quad\quad A = \frac{1}{R_{11}\tau'_{11}+1}$$
$$R_{15} = \frac{I_{11}}{h\nu_{11}}\sigma_{em11} \quad\quad B = \frac{R_{11}\tau'_{11}}{R_{11}\tau'_{11}+1}$$
$$\tau'_{11} = \frac{1}{\frac{1}{\tau_{11}}+R_{15}} \quad\quad C = \frac{W_{13}+R_{11}\frac{\tau'_{11}}{\tau_{11}}}{R_{11}\tau'_{11}+1}$$
$$W_{13} = \int i_{13,\nu}(\nu)\frac{1}{h\nu}\sigma_{abs13}(\nu)d\nu$$
$$W_{15} = \int i_{13,\nu}(\nu)\frac{1}{h\nu}\sigma_{ém13}(\nu)d\nu \quad D = \frac{W_{13}+R_{11}\frac{\tau'_{11}}{\tau_{11}}}{R_{11}\tau'_{11}+1}+\frac{1}{\tau_{13}}+W_{15}$$

(1)

the usual population equations for one site take the form given in (2).

$$n_{15} = A\times n_{tot} - A\times n_{13}$$
$$n_{11} = B\times n_{tot} - B\times n_{13} \quad\quad (2)$$
$$0 = C\times n_{tot} - D\times n_{13}$$

where "$n_{tot}$", is the total number of ions of the considered sites per volume. These equations are given for each site, then taking into account inhomogeneity. That means, we consider hundreds of systems like equation (2) in our model, each one at a different energy for the $^4I_{13/2}$ level. If we index these sites with k from 1 to K, then, after calculating the population inversion for each site with equation (2) and feeding equation (3) with the results, the three terms for emission (spontaneous for signal and stimulated for pump and signal) and absorption can be expressed under the form of (3):

$$di_{spont,\nu}(\nu) = \sum_{k=1}^{K} n_{13,k} \times \frac{g_k(\nu)}{\tau_{13,k}} \times h\nu \times dz$$
$$dI_{abs11} = \sum_{k=1}^{K}\left(-I_{11}\times\sigma_{abs11,k}(\nu_{11})\times n_{15,k} + I_{11}\times\sigma_{ém11,k}(\nu_{11})\times n_{11,k}\right)\times dz \quad (3)$$
$$di_{stimu13,\nu}(\nu) = \sum_{k=1}^{K} i_{13,k,\nu}(\nu)\times\left(n_{13,k}\times\sigma_{ém,k}(\nu)-n_{15,k}\times\sigma_{abs13,k}(\nu)\right)\times dz$$

"i" being the light power density per frequency unit and $g_k$ the normalized line profile of the spontaneous emission of the site number k (the $g_k$, as the one for the cross sections, are the usual lorentzian line shape for each site, but with a center varying inside the inhomogeneous line). We can now express the amount of light emitted by a bulk sample depending on the excitation parameters.

Because of the complexity of this modeling, fiber propagation is still in progress, but first results of the model on a bulk sample are still interesting because power density inputs in the model can be as large as the one reached in the fiber core.

In the following, we present the spectroscopic modeling for both simulated RFLN and GSHB results on a bulk sample, without propagation. In [12] we have shown that the homogeneous linewidth is far smaller than the inhomogeneous one at 77K in the same material. For both experiments modeling, we decided arbitrary but close to the reality to choose a ~20 ratio between homogeneous and inhomogeneous linewidth and to distribute 50 equidistant sites on each homogeneous line, one which allows modeling the inhomogeneous behavior in a reasonable time delay with a sufficient statistic.

To summarize we consider a homogeneous linewidth of 1.4 cm$^{-1}$(0.33 nm) given by our experiments and the literature, and an inhomogeneous line from 6517 to 6550 cm$^{-1}$ (1534.4 to 1526.7 nm) sampled by 1000 sites, corresponding to one site every 0,023 cm$^{-1}$ / (0.005 nm).

For RFLN simulation, the material is excited with very narrow (0.0023 cm$^{-1}$ / 0.0005 nm) laser lines, for different power densities. The emitted intensity is then calculated.

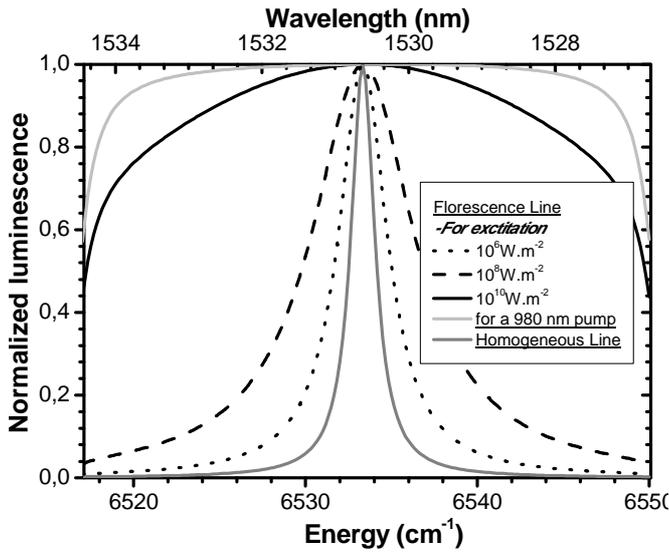

Fig. 10 : Normalized fluorescence intensity for different excitation powers (see text for description)

Fig. 10 presents the normalized fluorescence intensity for different excitation powers at 6534.6 cm$^{-1}$ (1530.3 nm): dotted line for $10^6$ W.m$^{-2}$, dashed line for $10^8$ W.m$^{-2}$, black solid line for $10^{10}$ W.m$^{-2}$. The gray line is the overall emission under a broad 980 nm (10204 cm$^{-1}$) excitation, and the dark gray line is the homogeneous line. In this figure, it is observed that the more the power density is large, the more the line width is broad. For a better understanding we have plotted the FWHM divided by the homogeneous line width as well as the peak intensity of the emitted light versus the excitation power density in Fig. 11.

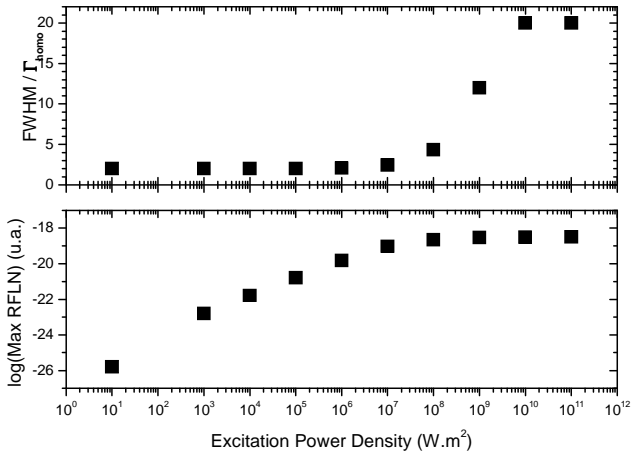

Fig. 11: Simulated RFLN FWHM divided by the homogeneous line width and the peak intensity of the emitted light versus the excitation power density

In Fig. 11, the FWHM begins to increase at a density of $10^6$ W.m$^{-2}$ when the fluorescence intensity reaches its maximum. Then the FWHM increases to a value of about 20 homogeneous line widths, that is the total width of the emission of the sample considered in the model.

In the GSHB simulation, the "simulated" sample was excited with $10^{15}$ W.m$^{-2}$ broad pump laser at 980 nm (10204 cm$^{-1}$) to fully invert all erbium sites in the considered volume. Then, a hole is burnt with a narrow (0.0023 cm$^{-1}$ / 0.5 pm) laser beam at 6534.6 cm$^{-1}$ (1530.3 nm). Since we model a bulk sample, spontaneous emission plays the same role of ASE as in the experiments of Arellano et al. [23].

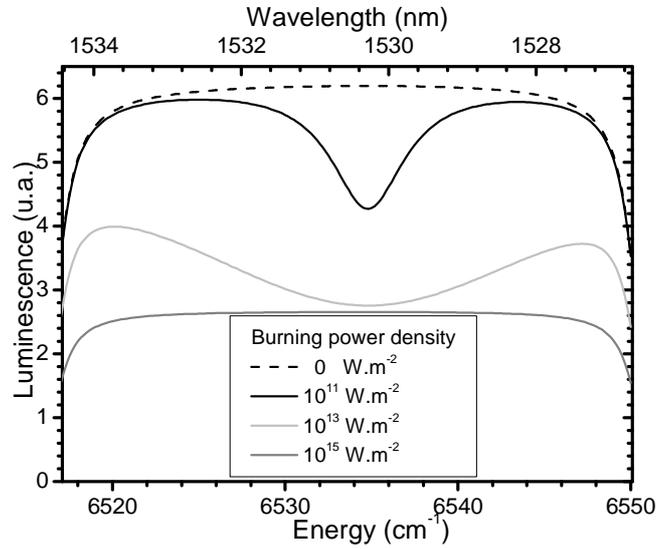

Fig. 12: Modeled emission spectra around 1530 nm for different burning power densities: 0 for dashed line, $10^{11}$ W.m-2 for black line, $10^{13}$ W.m$^{-2}$ for light gray line and $10^{15}$ W.m$^{-2}$ in dark gray line.

In Fig. 12, the inhomogeneous broad spectrum stands for emission without any burning power, and then a hole appears when a signal at 6534.6 cm$^{-1}$ (1530.3 nm) starts to saturate an electronic transition involving a class of sites. For a better understanding we have plotted the Full Width at Half Depth (FWHD) divided by the homogeneous line width and the relative hole depth intensity of the emitted light versus the burning power density in Fig. 13.

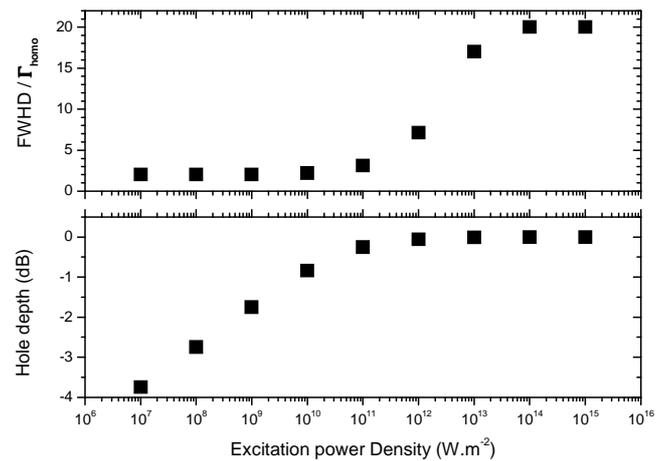

Fig. 13: Full width at half depth (FWHD) divided by the homogeneous line width and relative hole's depth intensity of the emitted light as a function of burning power density

In Fig. 13, the FWHD begins to increase at $10^{10}$ W.m$^{-2}$ when the hole depth reaches its maximum. Then the hole width reaches its maximum value that corresponds to the expected total width of the inhomogeneous spectral gain range.

## V. Discussion

In the experimental results reported in Fig. 3, the longitudinal RFLN spectrum exhibits a very broad profile attributed to an ASE process, while the transverse RFLN spectrum shows the usual luminescence spectrum for erbium in an aluminosilicate host (under the laser line). We have concluded that propagation of the luminescence signal modifies the spontaneous emission profile. We were not able to get enough RFLN signal from transverse RFLN measurement. For this reason, we chose to record pump power and length dependences of the longitudinal RFLN signal that indicate a limit of FWHM. In those experiments, the Lorentzien shape of the RFLN peak in Fig. 4 gave us a clue that this profile is interpreted as the homogeneous erbium emission line at the temperature of the experiment.

Moreover, even if we clearly record a power dependence, the value at low power is 2.8 cm$^{-1}$ (0.65 nm), which is the one found by Bigot *et al.* [12] in this material. To our knowledge, we report the first measurement of the homogeneous line width of the 1.5 µm erbium emission in a doped fiber by RFLN and this measurement is consistent with the literature.

In addition, we report GSHB measurement at the same temperature, showing strong power dependence too. At low excitation power, it is remarkable to get the same 2.8 cm$^{-1}$ (0.65 nm) value as for RFLN. If homogenous line width measurement by GSHB has been earlier reported [8, 24], its value was much larger than in our present investigation for both RFLN and GSHB experiments. Furthermore, we find the same 2.8 cm$^{-1}$ (0.65 nm) value of the homogeneous line width in aluminosilicate glass at 77 K by RFLN and by GSHB on fiber reported by Bigot *et al.* [12]. The discrepancy with the earlier works of GSHB [5, 6] was mostly due to the strong power dependence since it was not considered in the previous reports.

To exploit the experimental results of the line profile determination by our inhomogeneous simulation model we take advantage of the fact that the power density effect is more important on our RFLN experiments than the propagation effect. Our simulation in Fig. 10 shows that, in both RFLN and GSHB spectra, the line width at low power is two times the homogeneous line width as expected (the self-convolution of a lorentzian curve can be approximate by a lorentzian curve with the double width). However, when the power density increases, the line width increases up to the maximum value allowed by the simulation. In addition, we observe in Figs. 11 and 13 that, at low excitation power, the height of the peak (or the depth of the hole) strongly depends on power density, while, at high power, the height and the depth respectively reach an asymptotic value. The broadening and the height (and depth) of saturation take place at the same power density value. Indeed, these two effects are correlated to the inhomogeneous saturation of population of the $^4I_{13/2}$ level, for RFLN, and the $^4I_{15/2}$ level, for the GSHB.

In the RFLN experiment, the height of the peaks does not increase when all the erbium sites resonant with the excitation wavelength are still in their excited states (saturation of the absorption). Then the increase of power density will have no effect on the excited state population at the resonance wavelength. However, the erbium sites not exactly resonant with the excitation wavelength, contribute less to the intensity (lower cross section) than the ones exactly in resonance. These erbium sites absorb less light and saturation of absorption takes place at a higher excitation power. Indeed their emissions continue to grow after resonant saturation and make the line width broader. Then, when all the erbium sites population is saturated, both height and broadening of the line stop changing, as it is observed for very strong power density (Fig. 11).

The same behavior occurs in GSHB simulation (Fig. 13), but the energy levels are reversed. For the Fig. 12 the situation is slightly different. In fact, when the power of the burning signal is very high and begins to be comparable to the pump power, this signal will also play a pump role. Actually, the 3$^{rd}$ line from equation (3) shows that equilibrium between absorption and stimulated emission will occur. That is why, at very high power density, all the sites are saturated in this equilibrium and so the curve is flat.

Note that the same effects are observed on our experimental data, and even if the results are not quantitative, the orders of magnitude of power density when saturation takes place are similar in RFLN (100 µW in our fiber is ~10$^6$ W.m$^{-2}$). Besides, the behavior of GSHB is not similar because of the amplification effect which makes saturation depending of the 980 nm pump power that are really high in our model to ensure total population inversion.

However, this explanation is not consistent with the fact that power density broadening is more important at 1533 nm (6523 cm$^{-1}$) than at 1528 nm (6545 cm$^{-1}$). Indeed, propagation must be considered to give adequate explanation. Because absorption is more important at 1528 nm than at 1533 nm, the mean power propagating into the fiber will be larger at 1533 nm than at 1528 nm. That is why the broadening is more efficient at 1533 nm for the same input power than at 1528 nm.

## VI. Conclusion

After measuring the same value for homogeneous line width of the $^4I_{15/2} \leftrightarrow {^4I_{13/2}}$ zero line transition at 77 K by two different spectroscopic techniques (RFLN and GSHB), we simulate the behavior of the measurements with excitation power by modeling inhomogeneity in our materials.

These measurements, modeling and interpretations allow an understanding of the discrepancy between homogeneous line width measurements with the two techniques in the literature [12, 24] most probably due to the power dependence that causes inhomogeneous saturation of the excited state population.

Then we propose a model of simulation for bulk inhomogeneous erbium-doped materials. This model simulates the power dependences of both RFLN and GSHB measurements as an inhomogeneous saturation effect.

To be more realistic and quantitative, we need now to take into account phenomena such as propagation, up-conversion, energy transfers and even more than one class of sites [25] in our model. With those modifications we should be able to reproduce quantitatively the observed phenomena and better understanding of variability of GSHB power dependence with wavelength.

GSHB has been observed on other rare-earth-doped glasses [10] and could occur for any amplification or laser systems. A better understanding of the phenomenon can upgrade performances of amplification or laser devices. Finally, in the case of EDFA, building more complex models, including up-conversion and energy transfers, could give an explanation for the unexplained part of the losses in the yield of the device, and could even permit to better understand the pair concept given by Delevaque *et al.* [26].

ACKNOWLEDGMENT

The authors are grateful to the reviewers for the help to improve the manuscript.

REFERENCES

[1] A. L. Schawlow and C. H. Townes, "Infrared and optical masers," *Phys. Rev.*, vol. 112, pp. 1940–1949, Dec. 1958.
[2] T. H. Maiman, "Stimulated optical radiation in ruby," *Nature*, vol. 187, pp. 493–494, Aug. 1960.
[3] A. Yariv and J. P. Gordon, "The laser," in *Proc. IEEE*, vol. 51, pp. 4–29, Jan. 1963.
[4] R. Mears, L. Reekie, J. Jauncey, and P. D.N., "High-gain rare-earth rare-earth-doped fiber amplifier at 1.54μm," in *Proc. OFC, Reno, Nevada*, 1987.
[5] E. Desurvire, *Erbium-doped fiber amplifiers: Principles and Applications*. Series in Telecommunications and Signal Processing, John Wiley & Sons Inc, 2002.
[6] M. Tachibana, R. I. Laming, P. R. Morkel, and D. N. Payne, "Gain cross saturation and spectral hole burning in wideband erbium-doped fiber amplifiers," *Opt. Lett.*, vol. 16, no. 19, p. 1499, 1991.
[7] E. Desurvire, J. Sulhoff, J. Zyskind, and J. Simpson, "Study of spectral dependence of gain saturation and effect of inhomogeneous broadening in erbium-doped aluminosilicate fiber amplifiers," *IEEE Photon. Technol. Lett*, vol. 2, p. 653, sept. 1990.
[8] J. Zyskind, E. Desurvire, J. Sulhoff, and D. D. Giovanni, "Determination of homogeneous linewidth by spectral gain hole-burning in an erbium-doped fiber amplifier with geo2:sio2 core," *IEEE Photon. Technol. Lett*, vol. 2, p. 869, Dec. 1990.
[9] J. W. Sulhoff, A. K. Srivastava, C. Wolf, Y. Sun, and J. L. Zyskind, "Spectral-hole burning in erbium-doped silica and fluoride fibers," *IEEE Photon. Technol. Lett*, vol. 9, p. 1578, Dec. 1997.
[10] F. Roy, D. Bayart, C. Heerdt, A. Le Sauze, and B. Pascal, "Spectral hole burning measurement thulium-doped fiber amplifiers," *Opt. Lett.*, vol. 27, p. 10, Jan. 2001.
[11] T. Kushida and E. Takushi, "Determination of homogeneous spectral widths by fluorescence line narrowing in $Ca(PO_3)_2:Eu^{3+}$," *Phys. Rev. B*, vol. 12, p. 824, Aug. 1975.
[12] L. Bigot, A.-M. Jurdyc, B. Jacquier, L. Gasca, and D. Bayart, "Resonant fluorescence line narrowing measurements in erbium-doped glasses for optical amplifiers," *Phys. Rev. B*, vol. 66, p. 214204, Dec. 2002.
[13] R. Peretti, A. Jurdyc, B. Jacquier, E. Burov, and L. Gasca, "Resonant fluorescence line narrowing and gain spectral hole burning in erbium-doped fiber amplifier," *J. Lumin*, vol. 128, no. 5-6, pp. 1010–1012, 2008.
[14] M. N. Shunsuke Ono, Setsuhisa Tanabe and E. Ishikawa, "Origin of multi-hole structure in gain spectrum of erbium-doped fiber amplifier," in *OSA Trends Opt. Photo*, 2005.
[15] S. Ono, S. Tanabe, M. Nishihara, and E. Ishikawa, "Study on the dynamics of a gain spectral hole in a silica-based erbium-doped fiber at 77k," *JOSA B*, vol. 22, p. 1594, august 2005.
[16] T. Miyakawa and D. L. Dexter, "Phonon sidebands, multiphonon relaxation of excited states, and phonon-assisted energy transfer between ions in solids," *Phys. Rev. B*, vol. 1, pp. 2961–2969, Apr. 1970.
[17] F. Auzel, "Upconversion processes in coupled ion systems," *J. Lumin*, vol. 45, no. 1-6, pp. 341–345, 1990.
[18] F. Auzel, D. Meichenin, F. Pelle, and P. Goldner, "Cooperative luminescence as a defining process for re-ions clustering in glasses and crystals," *Opt. Mater*, vol. 4, pp. 35–41, Dec. 1994.
[19] C. R. Giles, E. Desurvire, and J. R. Simpson, "Transient gain and cross talk in erbium-doped fiber amplifiers," *Opt. Lett.*, vol. 14, no. 16, p. 880, 1989.
[20] C. Giles and E. Desurvire, "Modeling erbium-doped fiber amplifiers," *Journal of Lightwave Technology*, vol. 9, pp. 271–283, 1991.
[21] R. Laming, L. Reekie, P. Morkel, and D. Payne, "Multichannel crosstalk and pump noise characterisation of $Er^{3+}$-doped fibre amplifier pumped at 980 nm," *Electronics Letters*, vol. 25, no. 7, pp. 455–456, 1989.
[22] M. Yadlowsky and L. Button, "Pump-mediated inhomogeneous effects in EDFAs and their impact ongain spectral modeling," *Optical Fiber Communication Conference and Exhibit, 1998. OFC '98, Technical Digest*, vol. -, pp. 35–36, 1998.
[23] W. A. Arellano, M. Berendt, A. A. Rieznik, I. de Faria, and H. L. Fragnito, "Observation of spectral hole burning in the amplified spontaneous emission spectrum of erbium doped fibers," *IX Simpósio Brasileiro de Microondas e Optoeletrônica*, vol. 1, p. 1, 2000.
[24] E. Desurvire, J. Zyskind, and J. simpson, "spectral gain hole-burning at 1.53μm in erbium-doped fiber amplifiers *IEEE Photon. Technol. Lett*, vol. 2, p. 246, April 1990.
[25] R. Peretti, A.M Jurdyc, B. Jacquier, E. Burov, and A. Pastouret, "Evidence of two erbium sites in standard aluminosilicate glass for EDFA," *Opt. Express*, vol. 18, pp. 20661–20666, Sep. 2010.
[26] E. Delevaque, T. Georges, M. Monerie, P. Lamouler, and J.-F. Bayon, "Modeling of pair-induced quenching in erbium-doped silicate fibers," *IEEE Photon. Technol. Lett*, vol. 5, no. 1, pp. 73–75, 1993.

**Romain PERETTI** was born in Paris, France, on July 28, 1981. He received the M.S. degree in physics and the Diplome d'ingenieur in optics and photonics in 2005 1981, from respectively the University of Paris XI and Institut d'optique graduate school, France. In 2008 he received PhD in physics from the University of Lyon I, France. His doctoral work, done at Laboratoire de Physico-Chimie des Matériaux Luminescents (LPCML) in Villeurbanne in collaboration with Alcatel and Draka companies, involved Gain Spectral Hole Burning in EDFA's and nanostructuration of EDFA's.

From 2008 to 2009 he was a-Postdoctoral Research Affiliate at the LPCML working on Photodarkening of ytterbium doped fiber. He is now a-Postdoctoral Research Affiliate at Institut de Nanotechnology de Lyon (INL) involved in photonic crystal design and elaboration.

**Bernard Jacquier** received his PhD in Physics in 1975 from the University of Lyon in France. He joins the LPCML (Laboratoire de Physico-Chimie des Matériaux Luminescents) at the University of Lyon in 1969 and the CNRS in 1973. He has been Directeur de Recherche at CNRS since 1988. He spent many times abroad at Purdue University, Carleton University at Otawa, IBM Alamden, Montana State University at Bozeman, Christchurch University in New Zealand… Since 1986, he develops a pioneering work on active wave guide with a strong collaboration with Alcatel and now Draka companies. His scientific contribution to the field of rare earth-doped optical fiber spectroscopy and amplification properties was recognized together with Emmanuel Desurvire by a national Grand prize of the French Academy of Sciences in 2007.

**David Boivin** graduated in optics engineering from SupElec (Metz) in 2002. He received Ph. D. degrees from the university of Franche-Comté (Besançon, 2005) and from the Georgia Institute of Technology (Atlanta, 2006) for works devoted to telecommunication systems. Since 2007, he has been working in Draka Communications on new fiber measurements and characterizations techniques

**Ekaterina Burov** graduated in materials engineering from the Moscow Institute of Steel and Alloys Russia in 1989. She received a Ph. D. degree from the university of Paris VII in 1998 for a work devoted to the high temperature behaviors of silica polymorphs. She then joined Alcatel in the fiber optic R&D unit. Since 2007, she has been working in Draka Communications on the materials for the new telecommunication fibers.

**Anne-Marie Jurdyc** received her PhD in Physics in 1991 from the University of Lyon in France. The same year she integrates the LPCML (Laboratoire de Physico-Chimie des Matériaux Luminescents) at the University of Lyon as Chargée de Recherche for the CNRS (Centre National de Recherche Scientifique). In 1992, she spent one year at Rutgers University, NJ, USA, working in the group of E. Snitzer. Her research is based on spectroscopic characterizations of rare earth ions for telecommunication applications with Alcatel and Draka.